\documentclass[preprint,12pt]{elsarticle}

\usepackage{natbib}
\usepackage{url}
\usepackage{amssymb}
\usepackage{tabularx}
\usepackage{booktabs}
\usepackage{amsmath}

\journal{arxiv}

\begin{document}

\begin{frontmatter}



\title{Skull stripping with purely synthetic data} 


\author[label1]{Jong Sung Park}
\author[label2]{Juhyung Ha}
\author[label4]{Siddhesh Thakur}
\author[label3]{Alexandra Badea}
\author[label4]{Spyridon Bakas}
\author[label1]{Eleftherios Garyfallidis} 

\affiliation[label1]{organization={Intelligent Systems Engineering, Indiana University},
            addressline={107 S Indiana Ave}, 
            city={Bloomington},
            postcode={47405}, 
            state={IN},
            country={United States}}

\affiliation[label2]{organization={Computer Science, Indiana University},
            addressline={107 S Indiana Ave}, 
            city={Bloomington},
            postcode={47405}, 
            state={IN},
            country={United States}}

\affiliation[label3]{organization={School of Medicine, Duke University},
            addressline={1427 Fitzpatrick Center, Box 90281}, 
            city={Durham},
            postcode={27708}, 
            state={NC},
            country={United States}}

\affiliation[label4]{organization={School of Medicine, Indiana University},
            addressline={HITS building, St. 3100, 410 West 10th Street}, 
            city={Indianapolis},
            postcode={46202}, 
            state={IN},
            country={United States}}

\begin{abstract}
While many skull stripping algorithms have been developed for multi-modal and multi-species cases, there is still a lack of a fundamentally generalizable approach. We present PUMBA(PUrely synthetic Multimodal/species invariant Brain extrAction), a strategy to train a model for brain extraction with no real brain images or labels. Our results show that even without any real images or anatomical priors, the model achieves comparable accuracy in multi-modal, multi-species and pathological cases. This work presents a new direction of research for any generalizable medical image segmentation task.
\end{abstract}


\end{frontmatter}



\section{Introduction}\label{sec1}
Accurate brain imaging analysis is crucial for a wide range of neuroimaging applications, including disease diagnosis, surgical planning, and neuroscientific research. One of the fundamental preprocessing steps in brain MRI analysis is the removal of non-brain structures, a process known as skull stripping. Effective skull stripping ensures that subsequent image analysis—such as tractography and functional connectivity studies—is not confounded by non-brain structures.

Over the years, numerous algorithms have been developed, ranging from traditional intensity-based and morphological approaches to more advanced deep learning-based techniques. Classical methods, such as BET~\cite{smith2002fast}, have suggested using intensity thresholding, classification, surface deformation and etc. While these methods work well in many cases, they often struggle with variations in image quality, intensity non-uniformity, and anatomical differences among subjects. These factors can increase the need for hyperparameter tuning, complicating preprocessing pipelines. More recent machine learning and deep learning models have demonstrated significant improvements in accuracy and generalizability. These methods leverage large and diverse datasets to learn robust features, reducing the need for manual parameter tuning and improving performance across diverse imaging protocols. Despite the advancements in the field, challenges remain. Unique brains, such as pathological brains, neonatal and non-human brains can create disruption due to the reliability of the methods on their training data. Furthermore, many deep learning approaches require large, well-annotated datasets.

When training deep learning for skulls stripping, or any medical image segmentation models, we often aim for generalizability. The method should not be constrained on the training dataset, but should work on multiple scanners and subjects without critical failures. The Segment Anything Model (SAM)~\cite{kirillov2023segment} suggested using billions of training data to create a widely generalizable segmentation model, resulting in a method that could segment any image into multiple compartments. However, since we do not provide any additional information about the image, it will not be able to segment out a region of interest perfectly in medical imaging without some context.

This idea of utilizing a large dataset is not new in the medical imaging field as well. Training on a single task but with multiple modalities of images have been proposed before~\cite{isensee2019automated, thakur2020brain}. Several works have tried using SAM as a zero-shot model and compared to a task specific model, obtaining comparable, and often better results when given a precise prompt.~\cite{mohapatra2023sam, roy2023sam} Works such as MedSAM~\cite{ma2024segment} uses a similar idea as SAM and manages to provide a model that can be generalized across organs by training on a multitude of datasets consisting of different modalities and tasks. Domain transfer has been used as a good way of populating an otherwise insufficient dataset for training~\cite{zoetmulder2022domain}, while generative models have simulated images to populate the training image set as well~\cite{thambawita2022singan}.

In the brain imaging field, heavy augmentation to provide the best performance with limited data has been experimented. \citet{yeh2023brain} has shown that adding modifications to a template, including but not limited to cropping, adding heavy noise and textures and transformations can push a model to learn from a single atlas for any species. The synthetic ideas~\cite{gopinath2024synthetic} instead create synthetic images with random intensities based on anatomic labels of the human brain, training a model with 'infinite' modalities. Thus, we can force the model to learn structural aspects of the image rather than intensity, creating a modality-agnostic model. Nevertheless, both are still constrained on the atlas or the labels that were used to create the dataset and cannot escape the limitations of the training images.

Our idea stems from here. What is the essential information that the model should gather from the real images? Our hypothesis is that if we can build a model with minimal assumptions of the target segmentation task, it can achieve the maximum generalizability(widely applicable to any modality, species and unique cases). Thus, in this paper, we propose PUMBA(PUrely synthetic Multimodal/species invariant Brain extrAction), which aims to achieve generalizability among modalities, pathologies and species. Our approach is evaluated on public datasets including IXI~\cite{ixi}, MINDS~\cite{hata2023multi}, CAMRI~\cite{ds002868:1.0.1, ds002870:1.0.1}, LPBA40~\cite{lpba} and TCGA~\cite{Pedano2016-el, Scarpace2016-jr}, and several animal data as well and compared against existing state-of-the-art methods to demonstrate its efficacy. We provide the code in \url{https://github.com/pjsjongsung/PUMBA}

\section{Related Work}\label{sec2}

\subsection{Skull stripping}\label{subsec1}

A variety of skull stripping algorithms have been proposed before deep learning was widely used. Brain Extraction Tool (BET)~\cite{smith2002fast}, provided in FSL (the FMRIB Software Library)~\cite{jenkinson2012fsl} is based on intensity thresholding and surface deformation. After calculating the center of mass using binary thresholding, the method starts from an initial sphere that mostly fit inside the target's brain. The tessellated surface of the sphere is deformed following a similar intensity rule and smoothness regulation. Since the all voxel segmentation problem is reduced to tessellated surface deformation, the model is very fast compared to other methods. Thus, despite the frequent errors~\cite{atkins2002difficulties} and the dependence on some hyper parameters, have been one of the most used methods to this day. Analysis of Functional NeuroImages (AFNI)~\cite{cox1996afni} provides 3dSkullStrip, claiming to improve BET by adding additional priors to reduce false positives, including the eyes and the ventricles. Freeurfer~\cite{freesurfer} provides a segmentation algorithm that depends on the intensity difference between tissue and connectivity within tissue. A crude prediction from this is later deformed to match the brain shape. ROBEX (robust, learning-based brain extraction system)~\cite{robex} utilizes a random forest classifier on the features of the image (e.g. gradients, gaussian gradients, etc) to create a mask, which as other methods, are later fit to the brain's surface. Brain Surface Extractor (BSE)~\cite{shattuck2002brainsuite} applies anisotropic diffusion filters to process the image followed by edge detection. After removing small segments using morphological erosion, the selected region is expanded to match the shape of the brain.

Patchwise comparison is also a popular approach in brain segmentation. BEaST~\cite{beast} compares a set of templates to the target image. The comparison is made on multiple resolutions, where the output of each resolution is used on a higher resolution until it reaches the original scale. This reduces the covered voxels, reducing runtime and does a better job of removing false positives. ALFA (accurate learning with few atlases)~\cite{alfa} instead registers neonatal brain templates to the target image. A few closest templates are chosen and processed together to create a segmentation map.

Deep Learning has been a very popular approach in the brain extraction and brain segmentation field. One can categorize their novelty based on two aspects: 1. Model architecture 2. Training dataset.

For general U-Net architecture, nn-UNet~\cite{isensee2021nnu} has built a practical and very useful general rule of creating a model for specific datasets. Other pipelines have been proposed to accommodate more recent architectures~\cite{U-Mamba, gong2025nnmamba}. More brain extraction specific model architectures have been proposed as well. EVAC+~\cite{park2024multi} proposed integrating Conditional Random Field as RNN layers~\cite{arnab2018conditional} by engineering them specific for skull stripping tasks for better details in boundary segmentation. \citet{kim2024rgu} proposes using Ghost Modules~\cite{han2020ghostnet} with a 2D network to achieve computational efficiency. \citet{moazami2023probabilistic} generates uses an axial slice of the brain as a condition to generate multiple similar slices and creates a probability map of the brain mask through them.

Recently, it was shown that training a model with multimodal data rather than single model can be beneficial to its generalizability and even accuracy~\cite{ma2024segment}. HD-BET~\cite{isensee2019automated} was one of the first methods to train a single model with more than 2 image modalities. The training dataset included brains with glioblastoma and healthy subjects. BrainMaGe~\cite{thakur2020brain}, while using a similar idea, was trained to be more specific to brains with diffuse glioma. Both were created to work with any of the 4 structural modalities they were trained with.

\citet{dalca2018anatomical} proposed creating synthetic MRI images from adult human brain's anatomical labels for semi-unsupervised segmentation approach, when limited to images without any labels. Synthstrip~\cite{hoopes2022synthstrip}, inspired by this and the multimodal approach, creates synthetic images with a wider range of intensities. The images were created by assigning random intensities and deformation, along with some artifacts to the each of the labels in the label map. By replacing a real image and label pair with synthetic image and label pair, their model is able to train on theoretically infinite number of modalities, with the ground truth labels. Building on this idea, Synthseg~\cite{billot2023synthseg} was introduced for brain segmentation tasks as well.

Skull stripping is not limited to the human brain. Non-deep learning methods have been known to be more generalizable between species~\cite{wang2014knowledge}. Several animal specific methods~\cite{zhong2024nbest, hsu2020automatic} have been proposed by training the model on a certain species. \citet{wang2021u} have used a model trained on human data as a base model for few shot training on monkey data. Template based approaches have been studied as well. \citet{lohmeier2019atlasbrex} proposes registering a species specific template with a known brain mask to an image. \citet{yeh2023brain} instead augments the templates to simulate various scenarios when imaging each species/modalities for one shot learning. Various public datasets containing animal data, sometimes with masks and templates have helped facilitate such non-human brain extraction approaches~\cite{milham2018open, ds002868:1.0.1, ds002870:1.0.1}.

\subsection{Generalized segmentation in medical imaging}\label{subsec2}

There is still much research to be done on generalizable segmentation. Nevertheless, there have been research suggesting a generalized model can be beneficial to a level where it performs better than a task-specific model~\cite{ma2024segment}, suggesting that exploration of this field is crucial for improving segmentation results. For zero-shot segmentation, SAM~\cite{kirillov2023segment} provides good ground for computer vision images. A great deal of approaches have emerged that utilizes this model, either directly using with prompt(zero-shot learning) or by tailoring (few shot learning, fine-tuning) their SAM model more on specific task of their dataset. While we discuss some of the models here, authors refer the readers to the review that summarizes the recent developments~\cite{zhang2024segment}.

For zero-shot learning, \citet{mohapatra2023sam} compares SAM with Brain Extraction Tool (BET)~\cite{smith2002fast} and have showed that SAM with automatic bounding boxes can outperform BET in complex cases. \citet{roy2023sam} also compares the capability of SAM on Abdominal CT organ segmentation, showing that its performance largely varies on the prompts provided. Some more works.

Unlike using a computer vision model as a zero-shot or pre-trained model, populating a medical imaging training dataset is also a direction that can be utilized to achieve desirable generalizability. MedSAM~\cite{ma2024segment} trains on images from a variety of body regions and modalities. Fine-tuned from the weights of the original SAM model, it is able to give a robust result to any 2D medical image slice. However, there can be multiple tasks for any type of image, and one cannot guarantee a segmentation map correct for one task is correct for another task (e.g. tumor detection and brain segmentation).

Domain generalization is a way of training a model to work on an unseen domain(or modality in our case). \citet{zoetmulder2022domain} shows that domain transfer, generating an image of a different domain from a source domain, can be a good workaround of simulating a larger dataset when training a segmentation model for a small dataset of images. Creating synthetic images~\cite{hoopes2022synthstrip, billot2023synthseg} can also fall into this category. \citet{rafi2024domain} provides a good review regarding the recent work on domain generalization.

Unsupervised segmentation is a less studied but a set of methods with high potential in generalizability. \citet{kascenas2022denoising} suggests to train a denoising model that treats anomalies as noise as well. On the inference step, the denoised output and the original input can be compared to specify regions of interest. \citet{luo2023unsupervised} trains a wide range of healthy images for reconstruction and calculates the anomaly regions when reconstructing a brain with anomaly (tumors, lesions, etc). \citet{omidi2024unsupervised} propose training only on adult brains but with augmentation closer to fetal brain MRI. Unfortunately, we have yet to recognize a fully unsupervised skull stripping method using Deep Learning that does not rely on any data manipulation as of now.

\section{Methods}\label{sec3}
We describe the base idea and concept of our method below. Exact parameters and hardware used in our method can be found in the supplementary section. A grid search on different parameters have been conducted for finding the current optimal parameters.

\subsection{Assumptions}\label{subsec3}
Our main research question is what kind of prior assumptions the training set needs in order to be able to understand skull stripping. If we can use minimal information to train, intuitively it will also achieve maximum generalizability. Synthseg~\cite{billot2023synthseg} and Synthstrip~\cite{hoopes2022synthstrip} creates inifinite modalities of human brain MRI by assigning random intensities and small deformations to known human anatomical labels. One can interpret this as removing the intensity/contrast assumptions from the training set, which would have been given by the modalities the model is trained with in other models. We show, however, that this is not the limit to the priors one can remove for segmentation.

We first concentrate on the shape of the brain. It is a well known fact that a mammalian brain, regarding of species, is a fully connected chunk of tissue. It resembles a somewhat ellipsoid structure, and wider around the axial plane. By using these, and only these information, we can remove the assumptions on existence of any complex structure that is limited to the human species or its close relatives. This is especially important when we want to extend the model beyond human brains.

Next, neighboring tissues will share similar signal intensity range. Any boundaries between the tissue types, which includes the boundary between the brain and non-brain region, needs varying but identifiable contrast for brain extraction to be possible. By limiting the intensity assumptions to this, we can free the model from fitting to a specific intensity pattern to figure out the segmentation map. This is very similar to the Synthetic approaches~\cite{gopinath2024synthetic}, pushing the model towards multi-modality.

Finally, the brain structure of mammals shares the fact that the axial axis (or z-axis in voxel coordinates) is shorter than other axes. Providing this information to the training data can simplify the task from segmentation of the brain rotated in random positions to the segmentation of the brain in voxel space.

In theory, these assumptions should be able to cover any modality, mammalian species and pathology. We discuss how this is translated to code below.

\subsection{Creating images}\label{subsec4}

\begin{figure}[ht]
\centering
\includegraphics[width=0.9\textwidth]{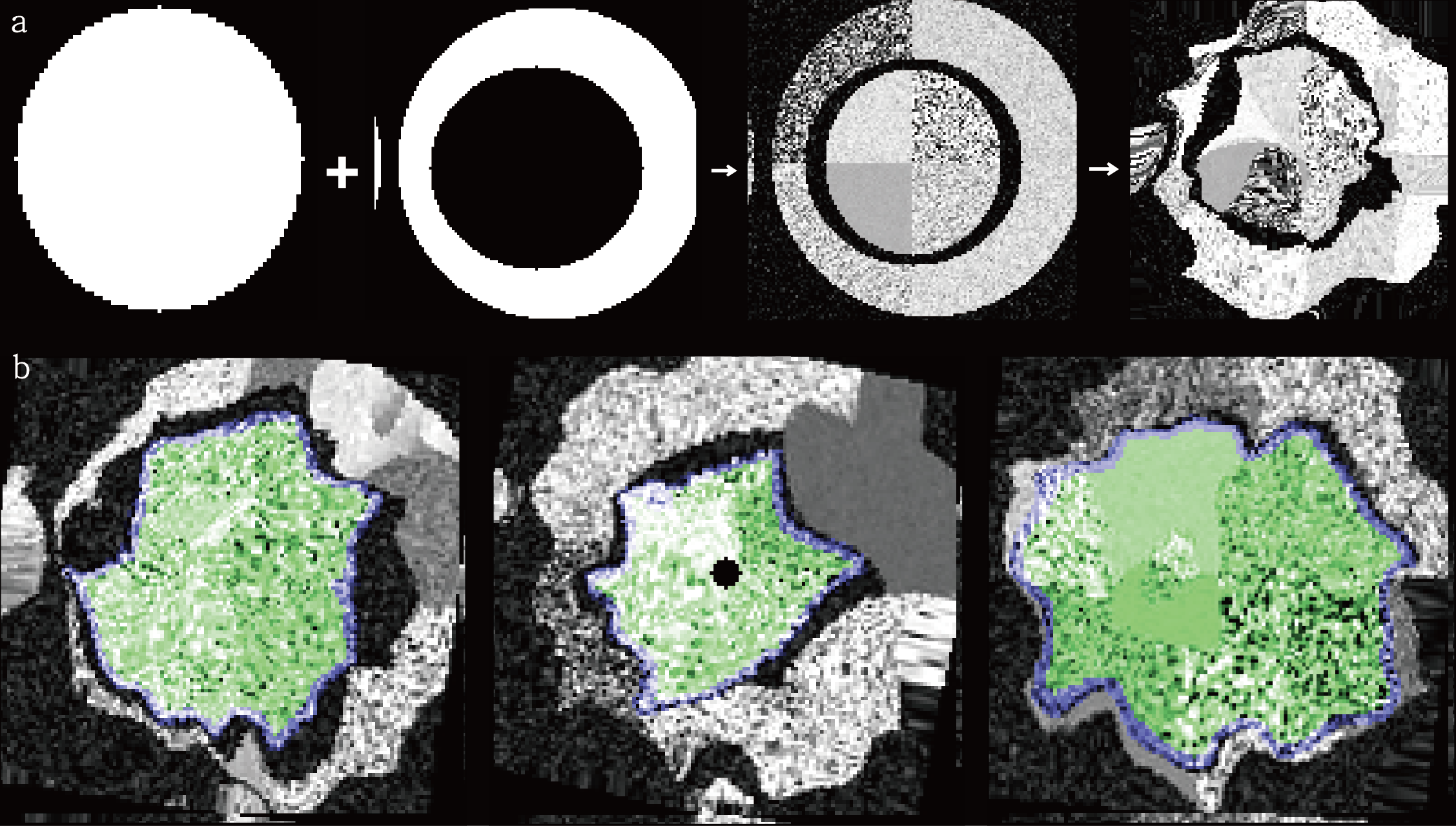}
\caption{A visual example of how the synthetic training data is generated. The steps of generating ellipsoids and applying random intensities and deformations is shown on a. Example images from the training dataset is shown on b. Green denotes the main mask, while blue denotes the boundaries.}\label{image_creation}
\end{figure}

\begin{figure}[ht]
\centering
\includegraphics[width=0.9\textwidth]{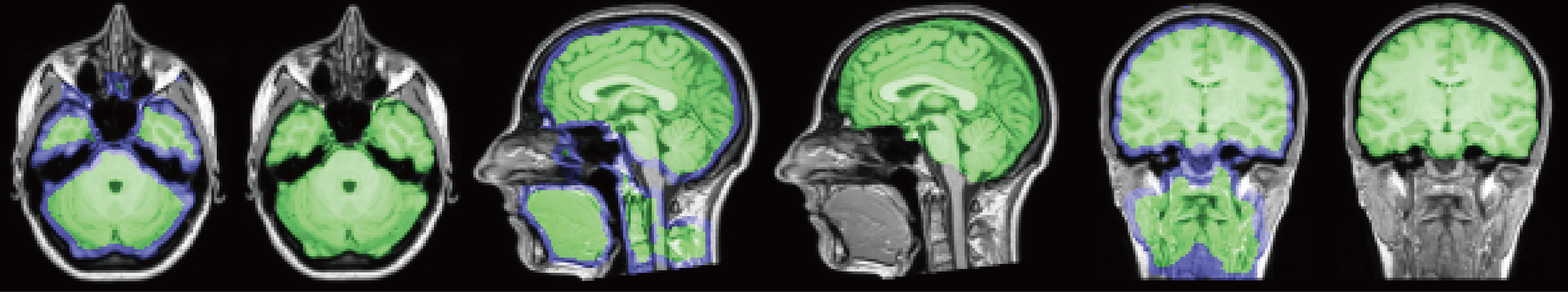}
\caption{The effect of the post processing step is shown on a T1 weighted image from the IXI dataset. Green (main mask) and blue region (boundary) is predicted from the model, which is processed to create the sole green mask.}\label{post_process}
\end{figure}

To simulate training images following the assumptions, we first create an ellipsoid with a random range on each axis. Then, using a random value of thickness, a hole the size of an ellipsoid is created in the initial ellipsoid. This creates a partially hollow shape, which is our simulation of the head. Another ellipsoid, again with a random range but smaller than the hollow space, is created inside. This represents the brain.

Further smaller ellipsoids and holes are added to simulate random artifact or anomalies. Afterwards, following the procedure suggested from \citet{hoopes2022synthstrip},  a smooth random deformation field is applied to the whole image to cover a variety of shapes that the ellipsoids can have. Note that all the steps here were applied on binary images(ellipsoids or background).

Next, we apply random intensities to all ellipsoids. To minimize the assumptions of intensity, we apply a gaussian noise with a set average and variance. Since both the brain and head can have multiple tissue types, instead of selecting a single mean and variance per ellipsoid, we divide the deformed ellipsoid to multiple areas, where each has a different gaussian patterns. Background is also added a gaussian signal with a relatively low mean and variance to train the model to be robust to noise. The smaller ellipsoids mentioned above is also given a random intensity but uniformly sampled from a smaller range. Fig.~\ref{image_creation} show the steps and example images of our method.

Finally, we create three labels: 1. brain, 2. brain boundary, and 3. background. Since the model might not have enough spatial information to separate the brain and non-brain tissues, other non-brain regions can be selected as regions of interest. The brain boundary serves as a method of selecting only the center most largest chunk, since it is mostly guaranteed that such shape is the brain in the image. A detailed example on how this changes the segmentation output can be seen in Fig.~\ref{post_process}.

\subsection{Training the model}\label{subsec6}
While it is theoretically best to create a new image each iteration of parameter updates, due to the time complexity we create a training dataset of size 3,000 beforehand. A simple U-net~\cite{ronneberger2015u} is created with 3D convolutions layers. Random flips, translations and rotations are done as augmentation steps. The code was written in Tensorflow~\cite{abadi2015tensorflow}.

\section{Results}\label{sec4}
To emphasize the generalizability of our approach, we quantitatively compare our method with BET~\cite{smith2002fast}, Synthstrip~\cite{hoopes2022synthstrip} and Unet-studio~\cite{yeh2023brain}. BET is currently the most generalizable method among traditional methods. Synthstrip, though created to target humans, is the most widely utilizable method that is a single model for all modalities. Unet-studio provides a separate single template trained model that could represent a more species/modality specific model. All methods are easy to run on multiple datasets, hence were chosen for this case.

\subsection{Quantitative results}\label{subsec7}

\begin{figure}[ht]
\centering
\includegraphics[width=0.9\textwidth]{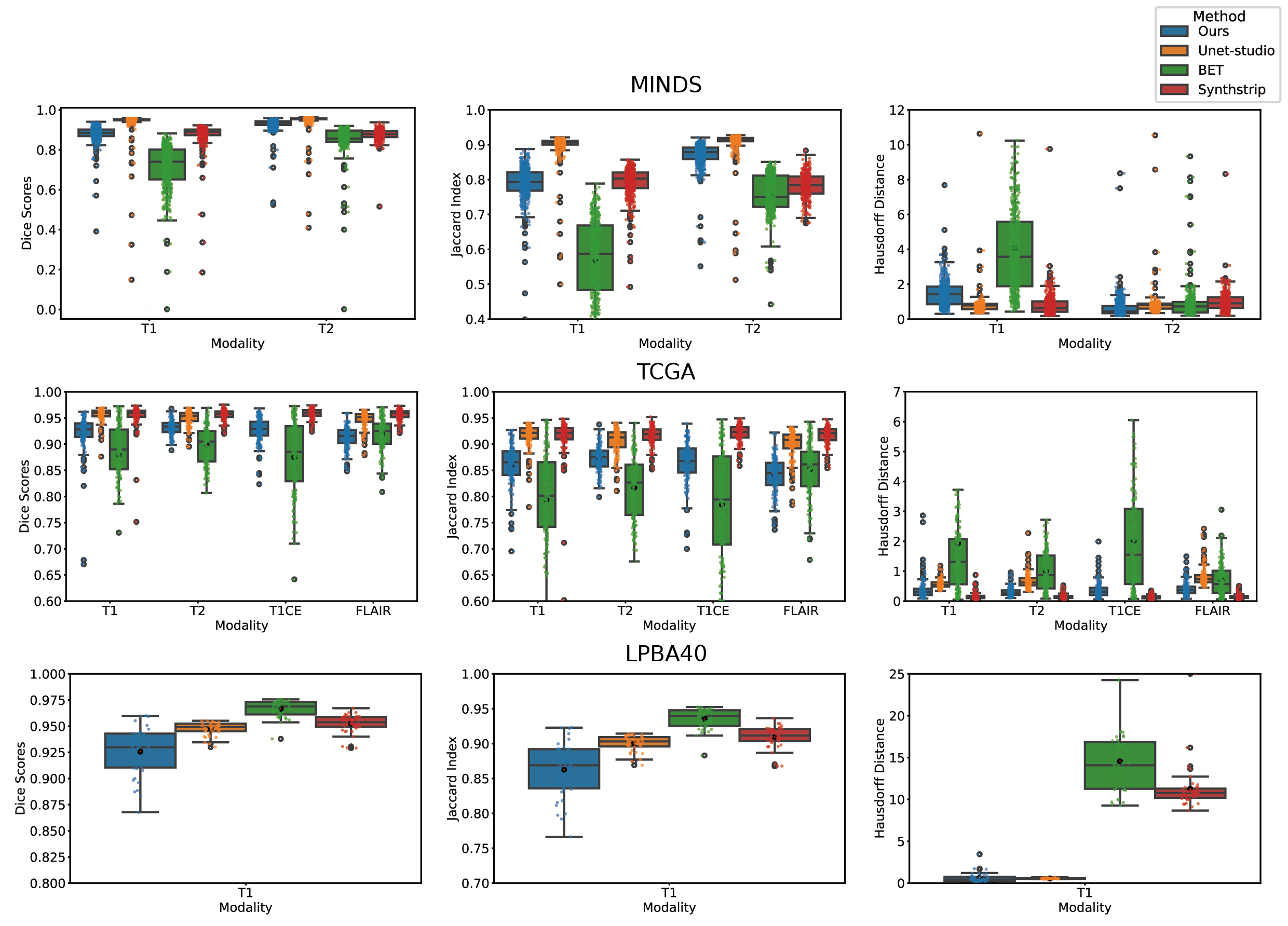}
\caption{The quantitative metrics are shown on the TCGA (top row), MINDS (middle row) and LPBA40 (bottom row) dataset. Dice Score (first column), Jaccard Index (second column) and Hausdorff Distance (third column) is shown. Note that unlike Dice Score and Jaccard Index, accurate segmentation results in smaller Hausdorff Distance.}\label{plots}
\end{figure}

In Fig.~\ref{plots} we evaluate our method using TCGA~\cite{Pedano2016-el, Scarpace2016-jr}, MINDS~\cite{hata2023multi} and LPBA40~\cite{lpba} dataset. Each dataset was selected to demonstrate pathology invariance, species invariance and performance on the most common case of brain extraction, on a human adult T1-weighted image. Our method shows comparable results, while showing the highest accuracy in marmoset data. It is important to mention that this metric was achievable without any image priors, supervised or unsupervised, when training the model.

\subsection{Qualitative results}\label{subsec8}

\begin{figure}[ht]
\centering
\includegraphics[width=0.9\textwidth]{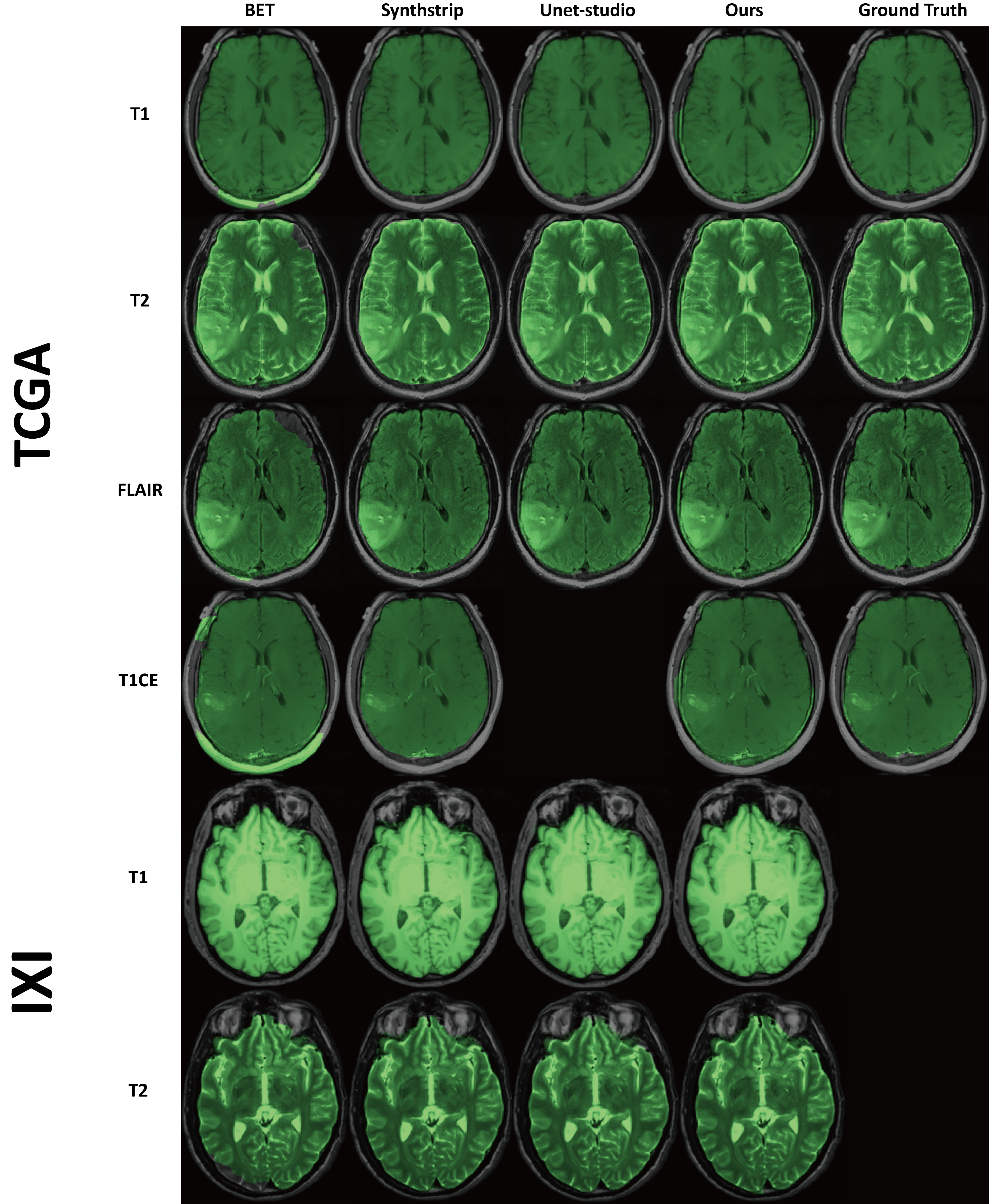}
\caption{The predicted mask of each method on multiple modalities of the TCGA (tumor) and the IXI (healthy) dataset. Ground truth is not provided for the IXI dataset.}\label{qual_human}
\end{figure}

\begin{figure}[ht]
\centering
\includegraphics[width=0.9\textwidth]{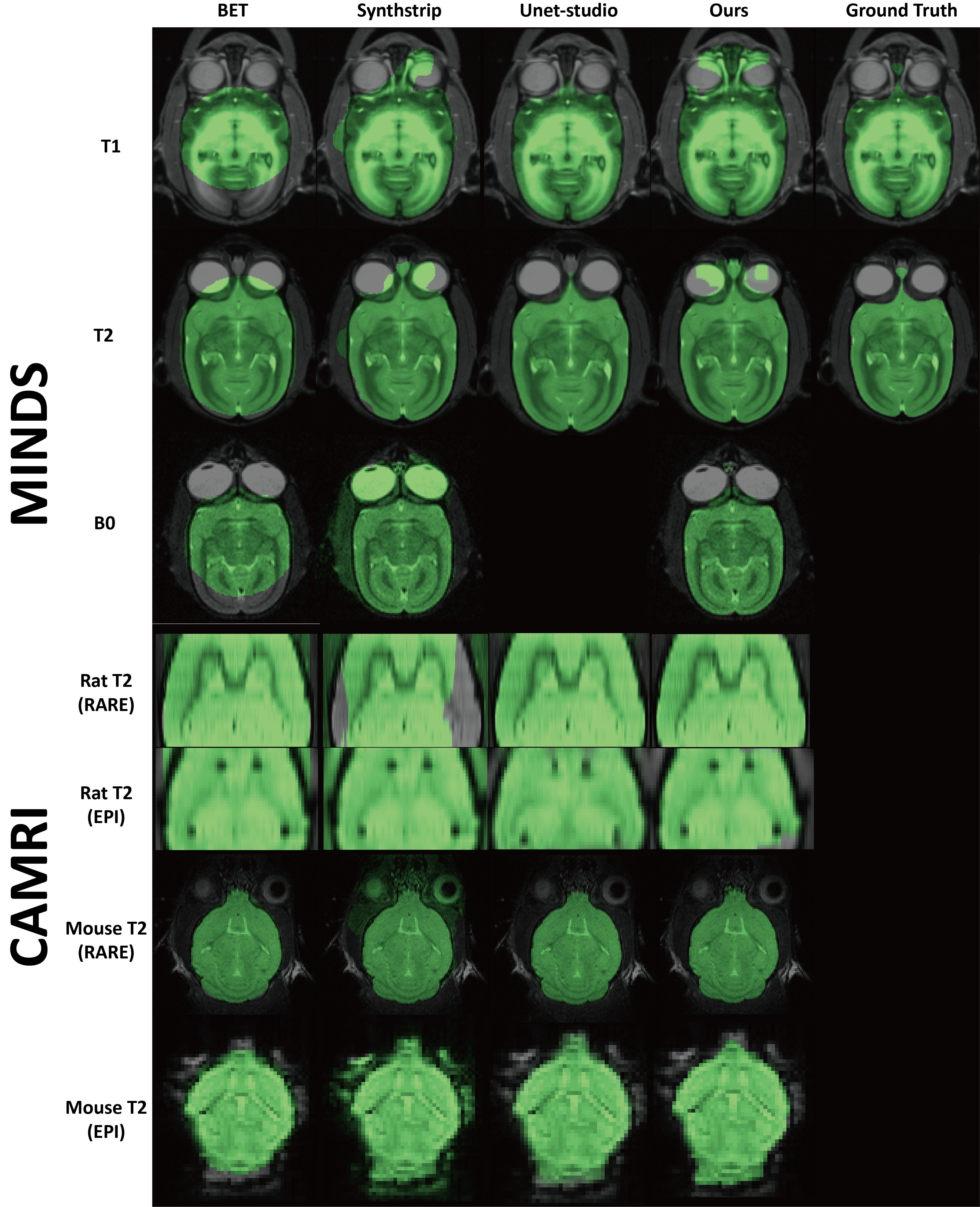}
\caption{The segmentation results on the MINDS (Marmoset) dataset and CAMRI (Rodent) dataset. Ground truth is shown where applicable.}\label{qual_animal1}
\end{figure}

\begin{figure}[ht]
\centering
\includegraphics[width=0.9\textwidth]{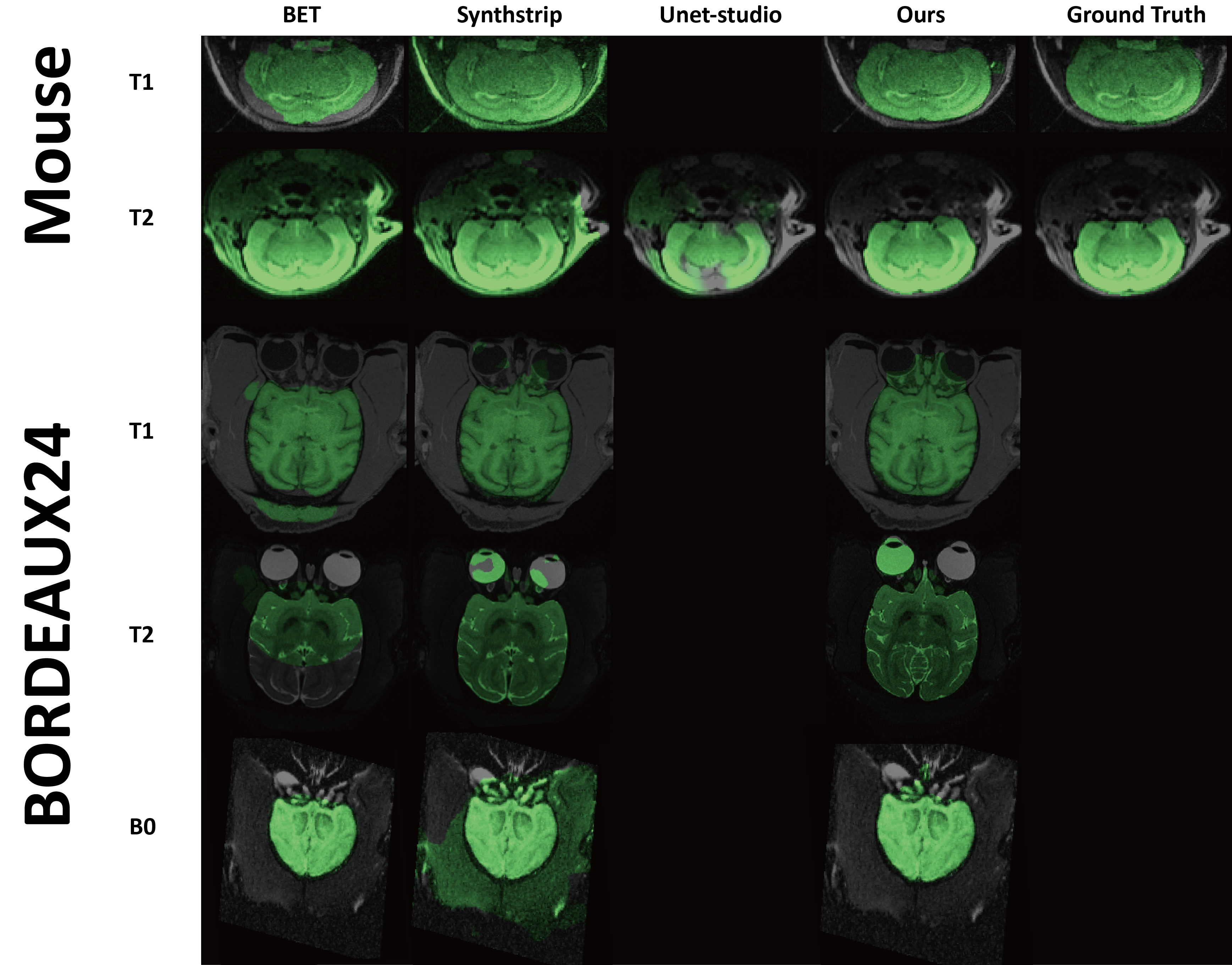}
\caption{The segmentation results on the privately gathered Mouse T1 and T2 images and BORDEAUX24 (Macaque) dataset. Ground truth is shown where applicable.}\label{qual_animal2}
\end{figure}

Fig.~\ref{qual_human}, Fig.~\ref{qual_animal1} and Fig.~\ref{qual_animal2} illustrates our model's capability on multiple type of images, including humans, control and pathological, marmosets, macaques and mice. Additional to the datasets from the previous section, we include IXI~\cite{ixi} for multi-modal human adult data, BORDEUX24 from the NHP dataset~\cite{milham2018open} for macaque brains, mouse and rat brains from the CAMRI dataset~\cite{ds002870:1.0.1} and individually gathered mouse brains. More explanation on the mouse data can be found in the appendix. The figure exhibits robust results of the predicted brain mask on all cases, whereas compared methods have a noticeable failure on images beyond the method was tuned for.

\section{Discussion}\label{sec5}
We present a completely non-atlas/non-label based synthetic training data generation for skull stripping. As far as we have acknowledged, this is the first attempt to train any segmentation model, not as a pre-training step but as a inference model, with synthetic data not biased by prior images.

It is worth noting that we have tried common fine-tuning approaches for few shot learning on more unique brains, such as rodents, to see the benefits of the model as a pre-trained model. However, we are yet to see any evidence that the feature extracted from this model can serve as a good starting point for training with a few images. Nevertheless, additional experiments with more tailored fine-tuning techniques to rule out its possibility.

As mentioned in the methods section, we have performed a grid search in finding the optimal parameters when generating the image. We have discovered that while most of the parameters have a noticeable affect on the model's performance, the largest difference is created by changing the range of intensities each ellipsoid can have. The best model was created from limiting both ellipsoid's intensity to be statistically higher than the background noise at a equal level. Future directions of the study could involve ways to set a more clear boundary between the background and the ellipsoids while introducing wider intensity range within the two ellipsoids to cover more possibilities of contrasts.

While the model's performance is impressive given the number of data it was trained on, we have observed instability when training the same pipeline with different images generated by the same rules. Despite the time complexity, to capture the true potential of the synthetic images, one should consider training the model online by creating a new image each iteration of training.

\section{Conclusion}\label{sec6}
We present PUMBA, a foundational model on completely synthetic unsupervised brain extraction. When deployed with different rules based on each specific cases, we believe the work can be extended beyond skull stripping to more unique medical image segmentation tasks, where an abundancy of ground truths or templates are often not available to train deep learning models.

\appendix
\section{Precised parameter information}\label{secA1}

\subsection{Image creation parameter}

We disclose the full intensity parameters in Table.~\ref{param_table1} that were used in creation of the synthetic images as a training data. Exact usage of the parameters and additional values used in deciding the shapes of the ellipsoids can be found in the code in our github repository.

\begin{table}[ht]
\caption{Intensity parameters when creating ellipsoids}\label{param_table1}%
\begin{tabular}{@{}ccccc@{}}
\toprule
 & Inner ellipsoid & Outer ellipsoid & Small ellipsoid & Background\\
\midrule
mean    & 0.4 $\sim$ 1.0   & 0.4 $\sim$ 1.0  & 1.0 & 0.1 \\
std   & 0.0 $\sim$ 0.4   & 0.0 $\sim$ 0.4  & 0.4 & 0.1 \\
\# of parts    & 4   & 4  & 1 & 1  \\
\bottomrule
\end{tabular}
\end{table}

\subsection{Model architecture}
We use a simple U-net~\cite{ronneberger2015u} architecture. Each encoder and decoder level contains two convolutional layers. Starting from the number of channels as 8, each level multiplies the number by 2. 5 different levels of scales were used in the model. Every input was transformed to a shape of (128, 128, 128) before running inference.

\section{Dataset information}\label{secA2}

The dataset used for training can be replicated by following the code in our github, \url{https://github.com/pjsjongsung/PUMBA}. The datasets used for testing\cite{ixi, lpba, milham2018open, ds002870:1.0.1, ds002868:1.0.1, Pedano2016-el, Scarpace2016-jr, hata2023multi} can all be found online and be downloaded publicly, with the exception of the privately collected mouse data. 

For the mouse data, T1W 3D RARE anatomical sequence were acquired with TR=250 ms, TE=25.7 ms, echo spacing 6.4 ms, RARE factor 8, 2 averages, 4 dummy scans, over 256x150x100 matrix, 25.6x15x10 mm FOV, reconstructed at 100 um resolution, and acquired in 15 minutes. T2W 2D RARE anatomical sequence were acquired with TR=2724 ms, TE=22 ms, echo spacing 7.3 ms, RARE factor 8, 8 averages, 12 dummy scans, comprising 30 interleaved slices with 250 um thickness and 300 um slice gap, and 128x128 in-plane matrix and 156.25 um isotropic resolution, producing 20x20x16.5 mm FOV, acquired in 4 minutes 21 seconds. All images were gathered using a 7T Bruker 70/20 system with an Avance III console and Paravsion 2.1.

\bibliographystyle{elsarticle-num-names} 
\bibliography{elsarticle-template-num-names}
\end{document}